\documentclass[conference]{IEEEtran}
\IEEEoverridecommandlockouts
\usepackage{cite}
\usepackage{amsmath,amssymb,amsfonts}
\usepackage{algorithmic}
\usepackage{graphicx}
\usepackage{subcaption}
\usepackage{textcomp}
\usepackage{xcolor}
\usepackage{setspace}
\usepackage[hyphens]{url}
\usepackage{hyperref}
\usepackage[hyphenbreaks]{breakurl}

\def\BibTeX{{\rm B\kern-.05em{\sc i\kern-.025em b}\kern-.08em
    T\kern-.1667em\lower.7ex\hbox{E}\kern-.125emX}}
\begin{document}

\title{Zero-shot Cross-lingual Voice Transfer for TTS}

\author{\IEEEauthorblockN{Fadi Biadsy*, Youzheng Chen*\thanks{* Equal Contribution}, Isaac Elias, Kyle Kastner,\\ Gary Wang, Andrew Rosenberg, Bhuvana Ramabhadran}
\IEEEauthorblockA{\textit{Google LLC} \\
\{biadsy,josephychen\}@google.com
}
}

\setstretch{0.96}

\maketitle

\begin{abstract}

In this paper, we introduce a zero-shot Voice Transfer (VT) module that can be seamlessly integrated into a multi-lingual Text-to-speech (TTS) system to transfer an individual's voice across languages. Our proposed VT module comprises a speaker-encoder that processes reference speech, a bottleneck layer, and residual adapters, connected to preexisting TTS layers. We compare the performance of various configurations of these components %
and report Mean Opinion Score (MOS) and Speaker Similarity across languages. Using a single English reference speech per speaker, we achieve an average voice transfer similarity score of 73\% across nine target languages. Vocal characteristics contribute significantly to the construction and perception of individual identity. The loss of one's voice, due to physical or neurological conditions, can lead to a profound sense of loss, impacting one's core identity.  As a case study, we demonstrate that our approach can not only transfer typical speech but also restore the voices of individuals with dysarthria, even when only atypical speech samples are available {\textemdash} a valuable utility for those who have never had typical speech or banked their voice. Cross-lingual typical audio samples, plus videos demonstrating voice restoration for dysarthric speakers are available here.~\cite{github}

\end{abstract}


\section{Introduction}

In recent years, there have been significant advances in Voice Transfer (VT) technology (e.g., \cite{li2024spontts, lee23i_interspeech, yuan2021improving}), integrated in Text-to-speech (TTS) (e.g., \cite{fujita2024lightweight}), Voice Conversion (VC) (e.g., \cite{yang22f_interspeech,lee23i_interspeech, chen21w_interspeech, doshi2021extending}), and Speech-to-speech Translation Models \cite{jia19_interspeech, jia2022translatotron, nachmani2024translatotron}. For example, Parrotron \cite{parrotron} is a VC model that converts atypical speech directly to a synthesized predetermined typical voice that can be more easily understood by others. Yet for many individuals with dysarthria, VT extends speech technologies to help them regain their original voice and potentially predict speech patterns they have lost or never had.

Recent research on TTS \cite{wang23c_interspeech,tran23d_interspeech} and VC \cite{yang22f_interspeech,lee23i_interspeech} have shown rapid progress on zero-shot or one-shot voice transfer and achieved great progress of speaker similarity on unseen speakers, but with the requirement of longer reference audio length \cite{wang23c_interspeech}, the cost of audio quality \cite{tran23d_interspeech}, or full fine-tuning \cite{yang22f_interspeech}. While Tran et al. \cite{tran23d_interspeech} describe work on cross-lingual voice transfer, this approach requires the language of reference audio to match the language of target audio.  The approach described in this work does not have such a requirement.

Voice characteristics are crucial to individual identity.
The loss of one's voice, caused by physical or neurological conditions, can result in a profound sense of loss, striking at the very heart of one's identity. Speakers with degenerative neural diseases, such as Amyotrophic Lateral Sclerosis (ALS), Parkinson's, and multiple sclerosis, may experience a degradation of some of the unique characteristics of their voice over time. Some individuals are born with conditions, like muscular dystrophy, that affect the articulatory system and limit their ability to produce certain sounds. Profound deafness also impacts vocal and articulatory patterns due to the absence of auditory input and feedback. These conditions present lifelong challenges in matching the typical speech heard widely.

In this paper, we describe a zero-shot VT module that can be easily plugged into a multi-lingual state-of-the-art TTS system \cite{saeki2024Virtuoso2}, using a single reference utterance. We demonstrate that such a module is capable of transferring voice across languages, even if the language of the input reference speech is different from the intended target language. %
Finally, we show that the same model produces high quality speech with high fidelity voice preservation even when the input reference is atypical, critical for those who have not banked their voice or never had typical speech (cf.~Section~\ref{sec:experiments:restore}).

The contributions of this paper are as follows: 
\begin{itemize}
\item 
We describe a zero-shot VT module that can be easily plugged into a state-of-the-art TTS system.
This module transfers voices given a single, short speech reference from each unseen speaker, with a high quality and fidelity.

\item
The proposed VT module is capable of transferring voice across languages, even when the language of the input reference speech is different from the target language.
\item
We introduce and compare novel bottleneck layers that have significant impact on zero-shot TTS quality and speaker similarity.
\item
We demonstrate that this model produces high quality speech with high fidelity voices across languages, even when the input reference is atypical, useful for those who have not banked their voice or never had typical speech. The reader is encouraged to listen to our audio and video samples \cite{github}.

\end{itemize}

The next Section ~\ref{sec:model} details our proposed voice transfer module, followed by ablation studies on the proposed architecture on the well-studied VCTK corpus~\cite{veaux2016superseded}. We discuss our cross-lingual results and voice restoration from atypical speech in Section~\ref{sec:experiments}.

\section{Model}
\label{sec:model}

\begin{figure}[ht]
\centering
\begin{subfigure}{\linewidth}
  \includegraphics[width=\linewidth]{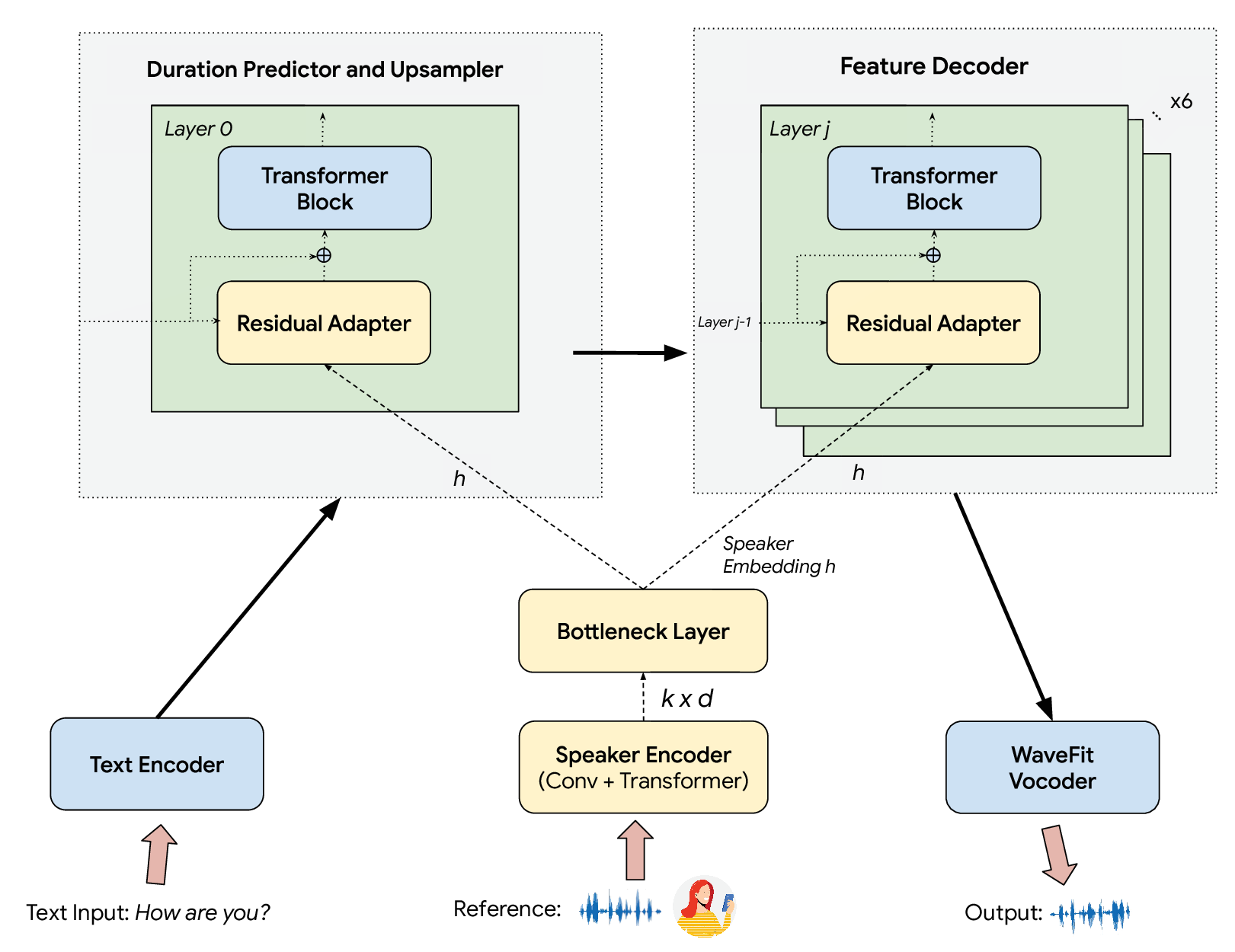}
  \caption{Model Architecture Overview.}
    \label{fig:ttsWithVT}
\end{subfigure}

\begin{subfigure}{\linewidth}
\centering
  \includegraphics[width=0.6\linewidth]{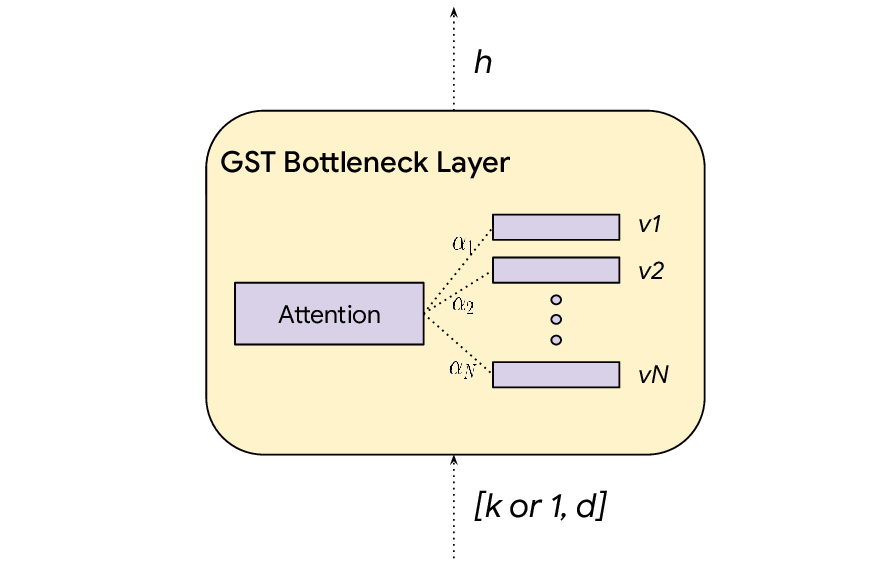}
  \caption{GST Bottleneck Layer.}
    \label{fig:gst}
\end{subfigure}
\caption{TTS Model Architecture with Voice Transfer.}
\vspace{-4mm}

\end{figure}

 The backbone multi-lingual TTS system used in this work and its training procedure are  described in detail in this previous work~\cite{saeki2024Virtuoso2}. This model is a joint speech-text model with feature-to-text (F2T) and text-to-feature (T2F) that are both jointly optimized on Automated Speech Recognition (ASR) and TTS data. It is trained with UTF-8 byte-based input representations derived from a text encoder which allows for sharing representations across languages efficiently.
The major components of the T2F model are based on Parallel Tacotron 2 \cite{elias2021parallel,elias21_ptaco2}. Inference begins with a text encoder that transforms the linguistic information into a sequence of hidden representations. These representations are then fed into a token duration predictor and upsampler, which generate a longer sequence of hidden representation proportional to the predicted output duration.  This expanded sequence is passed to a feature decoder to generate latent features. Finally, a WaveFit \cite{koizumi2023wavefit} vocoder converts these features into a  time-domain waveform output.  The inference flow through the model is shown in Figure~\ref{fig:ttsWithVT}. The yellow components represent the proposed VT module and related bottleneck layers (Figure~\ref{fig:gst}), discussed in the next section.

\subsection{VT Module}

 We extend this TTS system by adding a VT module that takes an input reference speech.  This extension enables the TTS model to transfer the voice in the reference speech to generate synthesized speech with this voice.
 
 The VT module is composed of (1) a speaker encoder that takes an 128-dimensional mel spectrogram from a reference utterance of $1$-$15$ seconds, to extract a high-level representation of this voice, using a stack of $5$ convolution layers with $3 \times 1$ filters followed by an $8$ Transformer layers, producing $1024$-dimensional embedding vectors. Pooling and L2 normalization of these hidden vectors, we construct an embedding tensor that summarizes the acoustic-phonetic and prosodic characteristics of the input reference utterance with one tensor.  This tensor is then passed to (2) a bottleneck layer to restrict the embedding space and to ensure its continuity and completeness. We find that the choice of the bottleneck has considerable impact on the Mean Opinion Score (MOS) and voice preservation. Finally, (3) the introduction of a residual adapter~\cite{rebuffi2017learning} between two consecutive layers in the duration and feature predictor blocks, as shown in Figure~\ref{fig:ttsWithVT}. The input to each residual adapter is a concatenation of the output of the bottleneck layer and the previous layer’s output, as shown in Figure~\ref{fig:ttsWithVT}.  The decision to use residual adapters is two-fold. First, it makes the model modular, i.e. one can enable or disable this component of the VT module without impacting the quality of the original TTS model, and train the VT module independently. Second, the parameters can be dynamically loaded on-the-fly.
 
 \subsection{Bottleneck Layers}
 \label{sec:bottlneck}
 
 In this Section, we conduct ablation studies on various types of bottleneck layers.  All bottleneck layers produce a summary $k\times 1024$-dimensional embedding tensor passed to the residual adapter in each layer, as described above. Depending on the bottleneck layer, however, the input to the bottleneck layer can be either a pooling of the sequence produced by the speaker encoder followed by an L2-normalization (i.e., $k=1$ in the Figure~\ref{fig:gst}) or a sequence of $k$-vectors passed to the bottleneck directly.  We describe the various types of bottleneck layers studied below.
 
  \textbf{VAE}: This comprises of a Variation Auto Encoder (VAE) layer using a Gaussian posterior probability distribution and a unit Gaussian prior. \cite{kingma2013auto} This bottleneck consumes the average pooling of the hidden representation sequence generated by the speaker encoder. At training time we used a KL-weight of 0.0001 and at inference time we used the posterior mode of the encoded references speech.

   \textbf{SharedGST}: This is a simplex-based bottleneck layer, employing a similar implementation of Global Style Token (GST) layer described in \cite{wang2018style}, using a $1024$ learned bank of vectors. This layer accepts a single pooled summary vector from the speaker encoder and it uses $4$-headed dot-product attention to compute its similarity to each of vectors in the the GST bank. The corresponding attention weights are then used to compute the weighted average of the vectors in the bank using the attention weights, to produce the final embedding vector, as shown in Figure~\ref{fig:gst}. This bottleneck  v
constrains the embedding vectors to lie within the learned simplex. Intuitively, one can view the vertices of the simplex as bases of voices, and a new voice is a convex combination of those voice bases. 

   \textbf{MultiGST}:  This bottleneck is similar to SharedGST with the same hyperparameters, except that we move the bottleneck layer and replicate it to each of the duration and feature predictor layers. In other words, we have a total of seven bottleneck layers (one bottleneck in the duration predictor and six bottleneck layers for the feature predictor). Each of these bottleneck layers consumes the same pooled vector from the speaker-encoder output. We hypothesize here that certain layers may benefit from learning a different simplex which allows the model to extract a different embedding information from the same speaker, benefiting each of those layers differently.

   \textbf{SegmentGST}: This bottleneck layer is similar to that of the Shared GST, utilizing the same hyperparameters. However, instead of pooling the speaker encoder output directly and feed it to the GST layer, we first allow the attention mechanism in GST to attend to the entire sequence from the speaker encoder. We do that, however, after reducing the sequence length by a factor of $16$, using $2$ $\times$ convolutional layers (each with $4$-strided $[8 \times 1]$ filters)  placed after the speaker encoder to locally extract information from wider context. Following the GST layer, we apply the average pooling before feeding the output to the residual adapters.

    It's worth noting that the intuition here is to shift from computing similarities to voice bases, as in SharedGST, towards calculating similarities to learned vertices, potentially representing acoustic-phonetic segments. By employing average pooling, we obtain centroids of these vectors, which serve as our embedding space. Importantly, the centroids of vectors within a simplex will also reside within the same simplex {\textemdash} a fundamental property of simplexes. We hypothesize that this embedding layer empowers the model to learn fine-grained speaker embedding vectors, as it attends to specific chunks from the original hidden representations from the reference, enabling a more nuanced representation of acoustic-phonetic information, plus we hypothesize it contributes to a smoother embedding space that that of SharedGST.

\section{Datasets and Model Training}
\label{sec:virtuoso2_datasets}

We follow a similar training recipe outlined in \cite{saeki2024Virtuoso2} to obtain a multi-lingual TTS system. In summary, the TTS system is trained within a joint speech-text training framework where both T2F and F2T are jointly optimized on ASR and TTS data. %
We utilize the features from a pre-trained conformer speech encoder, following a Universal Speech Model (USM) architecture as the T2F path. The USM was pretrained with BERT-based Speech pre-Training with Random projection Quantizer (BEST-RQ) and fine-tuned with the Multi-Objective Supervised pre-Training (MOST) objective~\cite{zhang2023USM,BestRQ}. We also make use of a pretrained WaveFit vocoder~\cite{koizumi2023wavefit} for the Feature-to-speech (F2S) component and keep it frozen.  Similar to~\cite{saeki2024Virtuoso2}, from the USM encoder, we use a split of $6$ of its Conformer blocks as the speech encoder and keep those frozen, and the rest of the $18$ USM Conformer blocks act as a shared encoder connected to an RNN-T decoder to perform the F2T path, effectively performing ASR training to provide alignments for the T2F path.

Since our extended model now accepts both text and reference speech, we pass a random consecutive chunk ($1$-$15$ seconds) from the target speech as a reference in each training sample. This helps prevent leakage of duration and linguistic information. The chunk length was sampled using a clipped Gaussian distribution with a mean of $8$ seconds and a standard deviation of $3$. We train the TTS model along with the VT module jointly on the multi-lingual training data described next.

The data is composed of large collection of transcribed, multi-lingual long-form YouTube data totalling around 200k hours and spanning several locales (ASR training data). %
The TTS data is composed of commercially licensed studio recordings, featuring 775 voice talents across several locales. %
The combined datasets cover 100+ locales.

\section{Experiments}
\label{sec:experiments}

We evaluate the effectiveness of our proposed VT module in performing zero-shot TTS across languages. We randomly select 25 speakers from the VCTK corpus validation set with reference utterances between $7$ and $14$ seconds. Note: VCTK is not been used in training any components of the model.  We generate 20 English sentences which are automatically translated into nine target languages: U.S. English, Arabic, Chinese Mandarin, French, German, Hindi, Italian, Japanese, and Spanish using a Large Language Model (LLM)~\cite{reid2024gemini}. The English sentences and their translations serve as the textual inputs for our TTS system in our experiments.
We conduct MOS naturalness and speaker similarity experiments for each of the nine languages across the four different bottleneck layers, discussed in Section~\ref{sec:bottlneck}. Each MOS experiment involves $500$ samples ($25$ speakers $\times$ $20$ sentences). Speaker similarity was measured by human raters. Raters were presented with pairs of audio samples: an English VCTK reference utterance and the utterance converted to one of the nine languages using the reference utterance, to determine if the pair seemed to have been spoken by the same speaker.

As shown in Table~\ref{tab:cross_lingual_vctk}, all bottleneck choices yield high-quality cross-lingual zero-shot voice transfer\footnote{Example audio samples are available at~\cite{github}}.  While the average MOS scores across languages are not significantly different for all bottlenecks, VAE and SegmentGST outperform the others in speaker similarity. 

Furthermore, we conducted Side-by-Side (SxS) experiments comparing these two bottlenecks across the nine languages. The results, with wins vs. losses between SegmentGST and VAE reported in the last column of the table, reveal a clear preference for SegmentGST. This preference is statistically significant in 7 out of 9 languages. Furthermore, two of these languages exhibit a significant preference across {\em all} approaches, while the rest show no statistical difference. Considering that SegmentGST demonstrates statistically superior MOS scores across languages and its similarity scores are not significantly worse, it emerges as a compelling choice for zero-shot voice transfer when the reference speech is typical speech.

\begin{table*}[t]
\footnotesize
    \centering
    \caption{Cross-lingual Zero-Shot Subjective Evaluations on VCTK Corpus.}
    \begin{tabular}{r|ll|ll|ll|lll}
        & \multicolumn{2}{c|}{VAE} & \multicolumn{2}{c|}{SharedGST} & \multicolumn{2}{c|}{MultiGST} & \multicolumn{3}{c}{SegmentGST} \\
        Language & MOS & Similarity & MOS & Similarity & MOS & Similarity & MOS & Similarity & SxS (wins/losses) \\
        \hline
        Reference & 3.3 $\pm$ .29 & 85\% $\pm$ 6\% & 3.3 $\pm$ .29 & 85\% $\pm$ 6\% & 3.3 $\pm$ .29 & 85\% $\pm$ 6\% & 3.3 $\pm$ .29 & 85\% $\pm$ 6\% & - \\
        \hline
        English & 3.7 $\pm$ .04 & \textbf{68\% $\pm$ 4\%} & 3.6 $\pm$ .05 & 48\% $\pm$ 4\% & 3.5 $\pm$ .05 & 58\% $\pm$ 4\% & \textbf{3.7 $\pm$ .05 } & 64\% $\pm$ 4\% & 335 / 265 $\dagger$ \\
        Chinese & 3.7 $\pm$ .08 & 67\% $\pm$ 14\% & 3.7 $\pm$ .06 & 70\% $\pm$ 14\% & \textbf{3.9 $\pm$ .05} & \textbf{70\% $\pm$ 14\%} & 3.9 $\pm$ .05 & 68\% $\pm$ 15\% & 407 / 193  $\dagger$ \\
        Spanish & 3.5 $\pm$ .05 & \textbf{72\% $\pm$ 7\%} & \textbf{3.8 $\pm$ .05} & 65\% $\pm$ 12\% & 3.6 $\pm$ .06 & 62\% $\pm$ 12\% & 3.6 $\pm$ .04 & 68\% $\pm$ 14\% & 507 / 93 $\dagger$ \\
        Arabic & 4.2 $\pm$ .03 & 90\% $\pm$ 7\% & 4.2 $\pm$ .03 & 90\% $\pm$ 7\% & \textbf{4.2 $\pm$ .03} & 74\% $\pm$ 11\% & 4.2 $\pm$ .03 & \textbf{90\% $\pm$ 6\%} & 308 / 292 \\
        French & 4.0 $\pm$ .03 & \textbf{88\% $\pm$ 5\%} & \textbf{4.3 $\pm$ .03} & 86\% $\pm$ 7\% & 4.1 $\pm$ .04 & 75\% $\pm$ 7\% & 4.0 $\pm$ .03 & 87\% $\pm$ 5\% & 363 / 237 $\dagger$ \\
        Japanese & \textbf{3.8 $\pm$ .05} & \textbf{77\% $\pm$ 5\%} & 3.4 $\pm$ .06 & 68\% $\pm$ 6\% & 3.6 $\pm$ .05 & 70\% $\pm$ 6\% & 3.7 $\pm$ .05 & 73\% $\pm$ 5\% & 320 / 280 \\
        German & 4.1 $\pm$ .04 & \textbf{85\% $\pm$ 6\%} & 4.1 $\pm$ .04 & 69\% $\pm$ 8\% & 4.0 $\pm$ .04 & 70\% $\pm$ 6\% & \textbf{4.1 $\pm$ .04} & 82\% $\pm$ 6\% & 347 / 253 $\dagger$ \\
        Italian & 3.6 $\pm$ .05 & \textbf{78\% $\pm$ 6\%} & 3.7 $\pm$ .05 & 72\% $\pm$ 7\% & 3.6 $\pm$ .05 & 70\% $\pm$ 6\% & \textbf{3.7 $\pm$ .04} & 76\% $\pm$ 6\% & 418 / 182 $\dagger$ $\ddagger$ \\
        Hindi & 3.9 $\pm$ .04 & \textbf{57\% $\pm$} 6\% & \textbf{4.1 $\pm$ .03} & 38\% $\pm$ 5\% & 4.1 $\pm$ .04 & 42\% $\pm$ 6\% & 4.1 $\pm$ .04 & 47\% $\pm$ 6\% & 391 / 209 $\dagger$ $\ddagger$ \\
        \hline
        Mean & 3.82 & \textbf{76\%} & 3.88 & 67\% & 3.85 & 66\% & \textbf{3.89} & 73\% & - \\
        Stddev & 0.25 & 11\% & 0.30 & 16\% & 0.26 & 11\% & 0.20 & 13\% & -\\
        \hline
        \multicolumn{1}{r}{Note:} & \multicolumn{9}{l}{$\dagger$~denotes SxS of SegmentGST significantly wins (p-value $<$ 0.01) against VAE;} \\
        \multicolumn{1}{r}{} & \multicolumn{9}{l}{$\ddagger$~denotes SxS of SegmentGST significantly wins against all other three setups.}
    \end{tabular}
    \label{tab:cross_lingual_vctk}
    \vspace{-5.5mm}
\end{table*}

\subsection{Model Behaviour on Dysarthric Reference Speech}

To assess the robustness of the proposed zero-shot TTS model with a broad range of atypical reference speakers, we selected 16 individuals with dysarthric speech of high severity from the Euphonia corpus~\cite{macdonald2021disordered}. Each of these speakers has one of the following etiologies: ALS, cerebral palsy, Ataxia, hearing impairment, vocal cord paralysis, or muscular dystrophy. The duration of every selected reference speech sample from each speaker is between $7-14$ seconds.

Utilizing the same 20 sentences as in previous experiments, we evaluated our proposed model with the four types of bottleneck layers%
using the atypical reference speech as input. Due to the sensitive nature of this data, we restrict our analysis to automatic ASR experiments on the synthesized output to approximate intelligibility. The average WER across the 16 speakers was:
$5.6\%$ for VAE; $2.7\%$ for SharedGST; $3.4\%$ for MultiGST; and $10.4\%$ for SegmentGST. According to paired T-test, we observed no significant difference between SharedGST and MultiGST. However, both of these bottleneck layers demonstrate significant improvements (p<0.01) over both VAE and SegmentGST, while there was no statistical significance between VAE and SegmentGST. 

Contrary to our findings with typical speech, we conclude that the highest quality zero-shot TTS for atypical reference speech can be achieved with shared and MultiGST bottleneck layers. Regrettably, we cannot assess speaker similarity in this context, as these speakers did not have the opportunity to bank their voices prior to the onset of voice degeneration.

\subsection{Case Study: Atypical Speech and Voice Restoration}
\label{sec:experiments:restore}

To demonstrate the system’s performance when atypical speech/voice is the only reference available and to study speaker similarity, we worked with two speakers, referred to as DK and AL. DK, who is profoundly deaf from a young age, has never had a typical voice. DK learned to speak English using Russian phonetics. Their speech patterns are unique and may be difficult for unfamiliar listeners to understand. We use $12$ seconds of DK’s atypical voice~\cite{github} as the reference speech and the associated ASR transcript~\cite{biadsy22_interspeech} and its translation into 7 languages (1 English plus 6 other target languages selected by DK) to demonstrate cross-lingual capability~\cite{github}.
Since DK can’t hear the synthesized speech, we asked 10 subjects who have working relationship with DK to score (1 to 10) the English output of how similar the voice is to that of DK. We observe that MultiGST scores the highest with an average of $8.1/10$ ($\pm 1.1$) followed by the VAE bottleneck with an average of $7$ ($\pm 1.4$).

The second speaker, AL suffers from muscular dystrophy, a condition that causes progressive muscle weakness and sometimes impacts speech production. Like DK, AL has never had typical speech. %
Similar to DK, we use 14 seconds of AL's atypical speech as reference and synthesize the ASR transcripts~\cite{biadsy22_interspeech} and its translations into 7 languages selected by AL~\cite{github}. We asked AL to evaluate how similar they think the output voice is to their own and they gave it $8/10$, preferring the generated output from that uses the SharedGST bottleneck layer.

\section{Reducing Misuse of VT for Atypical Speech}

We recognize that in the context of voice transfer technology, the potential for misuse of synthesized speech is a growing concern. To address this, we use audio watermarking~\cite{watermarking} to embed watermarks so synthesized speech from our model can be detected. This technique involves embedding imperceptible information within the synthesized audio waveform. This hidden data can be detected using specialized software, enabling the identification of potentially manipulated or misused audio content. It's important to note that the risk of misuse is significantly lower for individuals who have never had typical speech patterns. In such cases, the synthesized nature of the output would be readily apparent, minimizing the potential for deception.

\section{Conclusion}

We present a modular zero-shot VT module that can be easily incorporated into a preexisting multi-lingual TTS system, using residual adapters. We have discussed how this VT module can be trained jointly with the TTS system to obtain zero-shot cross-lingual capability, using a few seconds of reference speech. %
Our ablations show that the design and choice of bottleneck layer in the VT module have significant impact on voice quality and speaker fidelity. While all bottleneck configurations can perform well for cross-lingual zero-shot, SegmentGST obtain the highest MOS scores (an average of $3.9$) with high speaker similarity (an average of $73\%$), for typical speech reference across nine languages. This indicates that our raters, on average, $73\%$ of the time, perceived the speaker in the English reference speech to be the same speaker speaking the other eight languages. Finally, we show that SharedGST and MultiGST do remarkably well at restoring speaker voices when only atypical speech is available, achieving $~80\%$ speaker similarity in the two case studies, and a word error rate as low as $2.7\%$ across etiologies.

\section*{Acknowledgment}
The authors sincerely thank speakers DK and AL for their participation and numerous contributions to this work and their tireless support during the course of this research effort.

\bibliographystyle{IEEEtran}
\bibliography{refs}

\end{document}